\documentclass[sigconf]{acmart}
\usepackage{multirow}
\AtBeginDocument{%
  \providecommand\BibTeX{{%
    \normalfont B\kern-0.5em{\scshape i\kern-0.25em b}\kern-0.8em\TeX}}}

\copyrightyear{2021}
\acmYear{2021}
\setcopyright{rightsretained}
\acmConference[WSDM '21]{Proceedings of the Fourteenth ACM
International Conference on Web Search and Data Mining}{March 8--12,
2021}{Virtual Event, Israel}
\acmBooktitle{Proceedings of the Fourteenth ACM International
Conference on Web Search and Data Mining (WSDM '21), March 8--12,
2021, Virtual Event, Israel}\acmDOI{10.1145/3437963.3441714}
\acmISBN{978-1-4503-8297-7/21/03}
\vspace{-1em}
\begin{document}


\title{SoMin.ai: Personality-Driven Content Generation Platform}

\author{Aleksandr Farseev}
\email{farseev@itmo.ru}
\affiliation{%
  \institution{ITMO University}
  \streetaddress{Kronverksky Pr. 49}
  \city{St.Petersburg}
  \state{Russia}
  \postcode{197101}
}

\author{Qi Yang}
\email{yangqi@itmo.ru}
\affiliation{%
  \institution{ITMO University}
  \streetaddress{Kronverksky Pr. 49}
  \city{St.Petersburg}
  \state{Russia}
  \postcode{197101}
}

\author{Andrey Filchenkov}
\email{afilchenkov@itmo.ru}
\affiliation{%
  \institution{ITMO University}
  \streetaddress{Kronverksky Pr. 49}
  \city{St.Petersburg}
  \state{Russia}
  \postcode{197101}
}

\author{Kirill Lepikhin}
\email{kirill@somin.ai}
\affiliation{%
  \institution{SoMin.ai Research}
  \streetaddress{St. Petersburg, Medikov ave., 3, litera A.}
  \city{Russia}
  \postcode{197022}
}

\author{Yu-Yi Chu-Farseeva}
\email{joy@somin.ai}
\affiliation{%
  \institution{SoMin.ai Research}
  \streetaddress{71 Ayer Rajah Crescent}
  \city{Singapore}
  \postcode{139951}
}

\author{Daron-Benjamin Loo}
\email{elcdbl@nus.edu.sg}
\affiliation{%
  \institution{National University of Singapore}
  \streetaddress{21 Lower Kent Ridge Road}
  \city{Singapore}
  \postcode{119077}
}
\vspace{-1em}

\renewcommand{\shortauthors}{Farseev and Yang, et al.}

\begin{abstract}
In this technical demonstration, we showcase the World's first personality-driven marketing content generation platform, called SoMin.ai~\cite{farseev2018somin}. The platform combines deep multi-view personality profiling framework and style generative adversarial networks facilitating the automatic creation of content that appeals to different human personality types. The platform can be used for enhancement of the social networking user experience as well as for content marketing routines. Guided by the MBTI personality type, automatically derived from a user social network content, SoMin.ai generates new social media content based on the preferences of other users with a similar personality type aiming at enhancing the user experience on social networking venues as well diversifying the efforts of marketers when crafting new content for digital marketing campaigns. The real-time user feedback to the platform via the platform's GUI fine-tunes the content generation model and the evaluation results demonstrate the promising performance of the proposed multi-view personality profiling framework when being applied in the content generation scenario. By leveraging content generation at a large scale, marketers will be able to execute more effective digital marketing campaigns at a lower cost.
\end{abstract}

\maketitle
\pagestyle{plain}

\vspace{-1em}
\section{Introduction}
Driven by the rapid growth of Social Media in the past decade and, at the same time, by the drastic \textbf{decrease} of branded content organic reach, brands and digital media agencies are constantly boosting media budgets spent on digital advertising via programmatic platforms like Facebook Ads and Google Ads. Such an increasing scale of media buying activities raises the rapid demand on content validation and production technologies~\cite{gelly18,gelli2019learning} that are capable of easing the burden laying on the marketing teams when it comes to large-scale content creation and A/B testing.

Aiming to bridge such a technology gap, multiple multi-view data modelling~\cite{farseev2017tweetfit,farseev2015harvesting,samborskii2019whole} approaches have been proposed to solve the tasks of Content Popularity~\cite{gelly17,gelly18,gelli2019learning} and Engagement~\cite{ohsawa2013like, mazloom2016multimodal} Inference. Additional effort has been made to tackle the challenge of visual content generation, which has achieved significant success in Text Generation~\cite{chen2018adversarial}, Visual Arts Generation~\cite{lu2018image}, and Face Generation~\cite{karras2019style} domains. Unfortunately, the research efforts on the important space of Marketing Content Generation are relatively sparse, and there are not many AI models or industrial systems proposed to tackle this challenge. Even worse, the existing generative models are not tailored to fit the personal style of the target content viewer, reducing the impact of content personalization and, therefore, drastically diminishing the effectiveness of the content marketing as a whole.  The possible reason for the latter is the difficulty of adjusting image generation models to consider the marketing guidelines imposed by the brands as well as the overall diversity of marketing content styles and application scenarios across different industries. Simply put, modern content generation technology seems to be not yet capable in handling the complex business requirement of generating content at large scale, considering the behavioural preferences of the viewer, but also with no human intervention and presuming the compliance with brand marketing guidelines.

To take a further step towards achieving robust and reliable results of Marketing Content Generation, in this technical demonstration, we propose the World's first personality-driven content generation platform, called SoMin.ai~\cite{farseev2018somin}. The platform combines deep multi-view personality profiling framework and style generative adversarial networks for instant generation of branded marketing content that appeals to different human personality types. Guided by the MBTI personality type~\cite{buraya2018multi}, automatically derived from a user's social network content, SoMin.ai generates new social media content based on the preferences of other users with similar personality types aiming at enhancing the viewer experience when being exposed to digital Ads. Via the platform GUI, the demo user is also allowed to rate the generated content examples against the original branded marketing content and the feedback is used to guide the model in consequent content generation attempts.

\begin{figure}
    \vspace{-1em}
	\caption{SoMin.ai Platform Structure}
    \centering
	\includegraphics[width=0.40\textwidth]{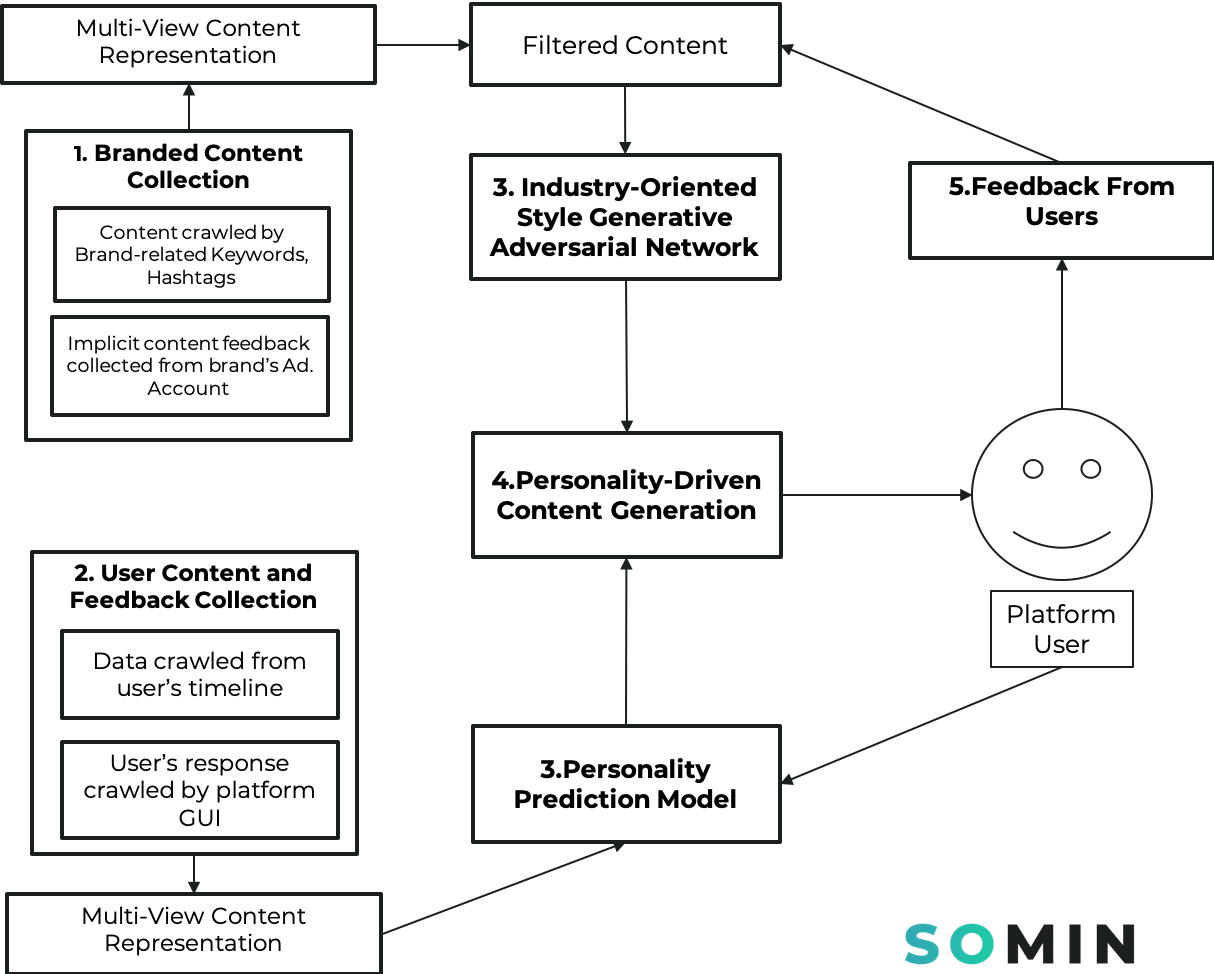}
	\label{fig:somin-structure}
	\vspace{-0.5em}
\end{figure}

\begin{figure}
    \vspace{-1.5em}
	\caption{Multi-View Personality Profiling Framework}
    \centering
	\includegraphics[width=0.40\textwidth]{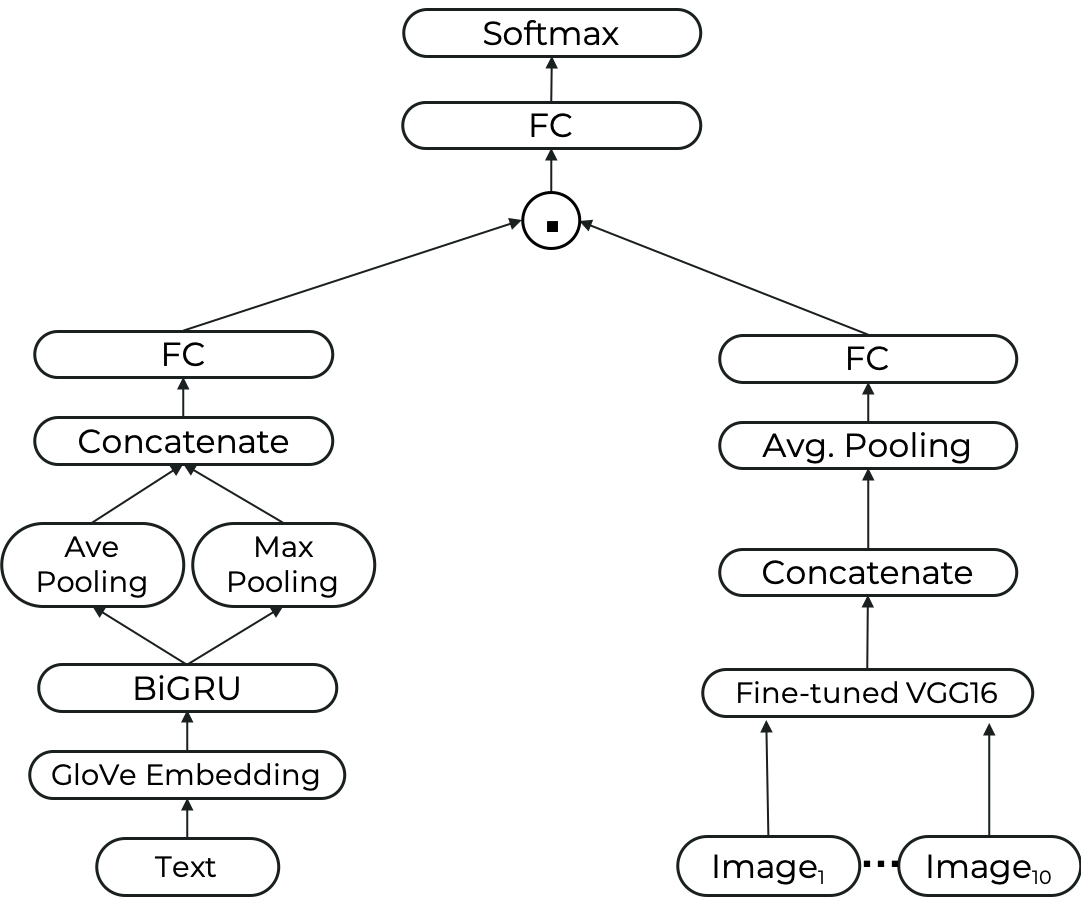}
	\label{fig:personality-frame}
	\vspace{-1em}
\end{figure}

\vspace{-0.4em}
\section{Platform description}
We now briefly describe the SoMin.ai platform. The structure of the platform is also displayed on Figure~\ref{fig:somin-structure}. From the figure, it can be seen that in the back-end, the branded marketing content gets collected via native interactions with brand Content Libraries and incorporated into the system every $12$ hours. The rest of the steps below are performed with different time intervals, ranging from $24$ hours to $30$ days: 

\noindent\textbf{1. Branded Content Collection:}
(A) Industry and Geographical Market-based brand content filtering via keyword search and visual content mining as search criteria. The step is necessary to narrow down model training data scope. (B) Implicit content response feedback collection from Brand Advertising accounts (e.g. clicks, likes, engagements).

\noindent\textbf{2. User Content and Feedback Collection:}
(A) Individual public social media content collection from Instagram and Twitter accounts of demo platform users for personality inference\footnote{The fixed-size multi-view dataset for training personality profiling model was previously collected and described by Yang et al.~\cite{qi2020know}}. (B) Demonstration user response collection via platform GUI.

\noindent\textbf{3. Generation and Profiling Models Training:}
(A) Filtered Past Marketing Content from Brands is used to train Industry-Specific style-based generator models~\cite{karras2019style}: one model for each selected industry (i.e. ''Automobile'', ''Fast Food'', and ''Fashion''). (B) Multi-View Public Social Media content is used to train Multi-View Fusion Personality profiling framework ''BiGRU + VGG''. The structure of the framework is outlined in Figure~\ref{fig:personality-frame}. From the figure, it can be seen that the social media post texts are first projected into a latent space via pre-trained embedding matrix and then fed into the bi-directional gated recurrent unit (BiGRU). The two output layers of the BiGRU then got concatenated and fed into a fully-connected layer (FC) to obtain the final textual data representation. Concurrently, fine-tuned VGG16~\cite{vgg} is used to extract visual features from the $10$ recent visual content and the average pooling layer is used to obtain the final image data representation. Finally, to obtain the latent multi-view representation and perform personality inference, the direct-product operation is applied and fed into the FC layer followed by the output prediction layer in the end.

\noindent\textbf{4. Personality-Driven Content Generation:}
(A) Personality type gets inferred by applying the prediction framework on recent individual public social media content from the demo user's timeline on Instagram or Twitter. (B) By applying cross-domain recommendation~\cite{farseev2017cross}, the best-performing content among users with similar MBTI personality types is retrieved from recent branded content for the chosen industry. (C) The content from the previous step gets fed into the industry-specific pre-trained style-based generator model for the generation of the new branded content versions. 

\noindent\textbf{5. Feedback Loop:}
Feedback from the Demo user was collected via the platform's GUI and incorporated into next platform runs. For example, if the platform has generated $5$ content variants ($Gen_1...Gen_5$) and the user has left feedback about Images $Gen_1$, $Gen_3$, $Gen_4$ and the original image $Org$ as her preference, then the source style images caused generations that led to unsuccessful results ($Gen_2$, $Gen_5$) will be ''penalized'' by excluding them from the next re-training cycle, and, at the same time, the successful source images, will be prioritized and diversified by Invariant Augmentation~\cite{shorten2019survey} during the next model re-training.

\vspace{-1.4em}
\section{The Demonstration}
\begin{figure}
    \vspace{-1em}
	\caption{SoMin.ai Online GUI Snapshot}
	\includegraphics[width=0.45\textwidth]{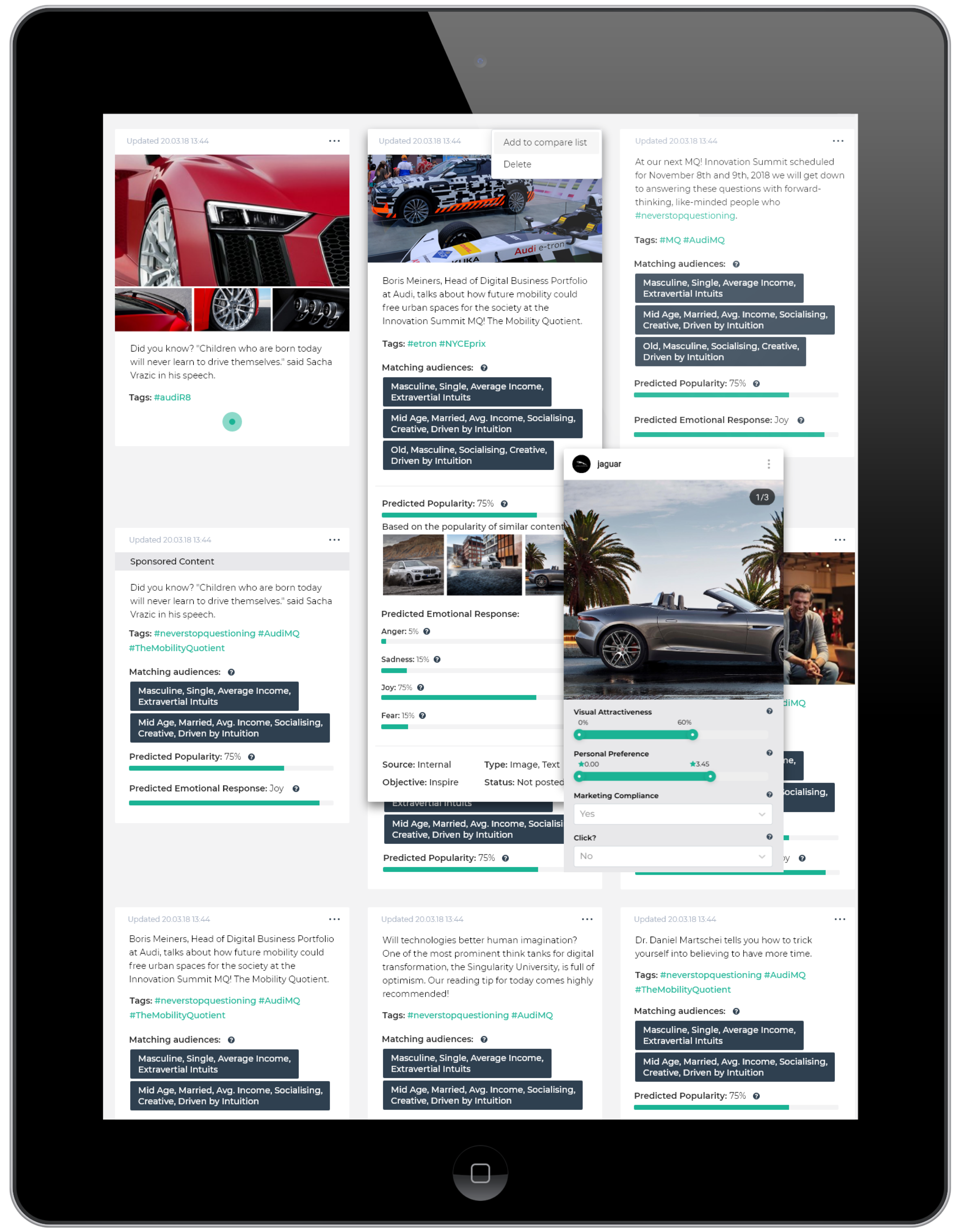}
	\label{fig:somin-ui}
	\vspace{-0.8em}
\end{figure}

Using our web GUI displayed on Figure~\ref{fig:somin-ui}, SoMin.ai users can specify their social media handle on Twitter or Instagram to instantly receive generated content examples according to their-predicted personality traits and preferences. The visual content generation response will be given in a form of marketing post previews, where each preview will be listed together with contextual information outlining the basis behind each particular generation: audiences, predicted popularity and viewer personality, emotional response, etc. The original Brand-crafted source content examples will be also organized in random positions on the dashboard for evaluation purposes. By clicking the content previews in the dashboard, the demo platform users will be routed to the ''online feedback'' form, where they will be able to indicate their preferences regarding the four key content marketing evaluation factors reflecting the content generation performance: a) content visual attractiveness (score $0-100\%$); b) content personal preference  (score $1-5$); c) industry and brand marketing compliance (''Yes'', ''No'', ''Do Not Know''); d) whether they would click the content if they see it as a banner Advertisement on Facebook (''Yes'', ''No''). Finally, the Demo users will be able to re-post the generated content that received the highest overall score on their Twitter timelines with a reference to the $14$th ACM International WSDM Conference and this technical demonstration track.

\begin{table*}[]
\addtolength{\tabcolsep}{0pt}
\caption{Personality Prediction Performance (Macro F1)}
\vspace{-1em}
\begin{tabular}{c|c|c|c|c|c|c|c|c|c|c|c|c}
\hline
\multirow{2}{*}{Model}          & \multicolumn{4}{c|}{Text} & \multicolumn{4}{c|}{Image} & \multicolumn{4}{c}{Text+Image}                      \\ \cline{2-13} 
                                & EI   & SN   & TF   & JP   & EI    & SN   & TF   & JP   & EI            & SN            & TF            & JP   \\ \hline
Logistic Regression             & 0.76 & 0.46 & 0.54 & 0.61 & 0.61  & 0.52 & 0.54 & 0.61 & 0.77          & 0.5           & 0.54          & 0.63 \\ \cline{1-1}
Support Vector Machine          & 0.77 & 0.46 & 0.55 & 0.61 & 0.6   & 0.52 & 0.55 & 0.61 & 0.79          & 0.5           & 0.55          & 0.61 \\ \cline{1-1}
Light Gradient Boosting Machine & 0.79 & 0.49 & 0.58 & 0.61 & 0.63  & 0.54 & 0.55 & 0.61 & 0.79          & 0.54          & 0.58          & 0.62 \\ \hline
Our Approach (BiGRU+VGG)~\cite{qi2020know}         & 0.81 & 0.56 & 0.62 & 0.59 & 0.66  & 0.52 & 0.58 & 0.59 & \textbf{0.83} & \textbf{0.57} & \textbf{0.64} & 0.6  \\ \hline
\end{tabular}
\label{tab:evaluation}
\vspace{-1em}
\end{table*}

\vspace{-1em}
\section{Performance Evaluation}
Except for the content generation performance evaluation that will be conducted based on the live user feedback and reflected in the platform GUI during the conference, we have additionally assessed the power of our personality prediction framework, called 'BiGRU+VGG''~\cite{qi2020know} over the $4$ key MBTI personality traits: Introversion/Extroversion (EI), Sensing/iNtuition (SN), Thinking/Feeling (TF), and Judging/Perceiving (JP). The evaluation results are presented in the Table~\ref{tab:evaluation}. From the table, it can be seen that the fusion model effectively solves the problem of personality prediction based on users' public multi-view data from social media as compared to the case when being trained on either textual or visual data independently. Moreover, it can be noticed that the best prediction performance was achieved for EI personality label, uncovering the strong bond between social media content and human personality traits. The latter observation also allows us to hypothesise that the above two human personality aspects might be of greater influence on the marketing content user preferences when it comes to viewing digital Ads on Social Media. Last but not least, we would like to highlight the overall promising performance achieved by the ''BiGRU+VGG'' framework, encouraging future research attempts along the personality profiling and marketing content generation directions.

\vspace{-1.5em}
\section{Target Users \& Commercial Impact}
It has been long known that Creative Content is one of the key crucial aspects defining the performance of Digital Ads along with both Organic~\cite{zarrella2009social} and Paid Media~\cite{keller2009role} digital marketing directions. High-quality content serves the basis of the modern Content Marketing~\cite{baltes2015content} paradigm, and often dictates advertisement campaign relevance score~\cite{wang2011relevance}, which, in turn, is directly related to the success of the Advertisement Campaigns in terms of both scale and cost-efficiency.

Modern programmatic advertisement delivery engines no longer require advertisers to rely just on targeting categories and a limited number of creative assets to run the Ads successfully, but rather are capable at handling thousands of content variations for learning which content needs to be shown to different kinds of audiences to maximize user engagement and conversion rates. The technology is called Dynamic Content/Creative Optimization (DCO) and widely considered to be the future of digital advertising~\cite{haberman2016dynamic, kulkarni2018dynamic, facebookdco20}.

Unfortunately, the two key digital advertiser groups (Brands and Digital Marketing Agencies) still have to rely on their intuition, tremendous amounts of manual work, and the ''gut feeling'' when approaching the task of creative content production. The amount of content that human teams are able to produce is almost never sufficient for holistic A/B testing~\cite{dixon2011b} and DCO~\cite{haberman2016dynamic} campaigns. At the same time, these technologies are considered to be prominent future development directions of the state-of-the-art Digital Advertisement platforms~\cite{haberman2016dynamic, facebookdco20}. Such a mismatch of human abilities and advertisement technology creates a huge Research and Industry gap resulting in a tremendous waste of digital advertisement budgets on inefficient and limited advertisement campaigns over a small number of likely ineffective creative content assets while under-utilizing the power of DCO capabilities.

This Technical Demonstration aims at addressing the above gap by introducing the world's first personality-driven content generation platform that is able to leverage on automatically-detected personality traits to generate marketing content for DCO campaigns and large-scale content A/B testing. The systems of this kind open an opportunity for brands and agencies to produce thousands of content variations based on one or a small number of input content examples to use such generated content in DCO campaigns for letting programmatic engines to leverage on their massive scale of accumulated user behaviour data and find the right audiences for the right creative assets.

During the past few years, several commercial AI systems have been proposed in an attempt to address the marketing content generation problem and, not surprisingly, immediately were surrounded with considerable attention from the key research and industrial players in the Digital Advertisement space. For example, an intelligent AI-Driven banner-generation technology was proposed by Alibaba in their product Alibaba Luban~\cite{liu2019intelligent}, while PUMA Asia Pacific has recently introduced the World's First AI-Generated Social Media Influencer, called Maya~\cite{techinasiaMaya}. The former technology is definitely a breakthrough in the industry, but currently only adopted to Alibaba's own e-commerce infrastructure in China and yet to be generalized to be applied outside Alibaba; the latter project was developed more along with appearance/face content generation direction and, therefore, might need to undergo a significant restructuring to be capable at solving the our-defined problem. Furthermore, several emerging tech startups have been aiming at solving content generation problems in other domains. For example, Persado.com and Datasine.com are aiming at improving Digital Advertisement performance via the automatic textual content generation and visual content recommendation, respectively. At the same time, SocialBakers.com offers a content recommendation tool, which helps to choose likely-performing content, but does not generate new creative assets to use in DCO campaigns. Again, as it can be seen, the above projects either solve the problems in their respective narrow domains or, alternatively, only help in choosing content from the existing assets but not in new content generation. Therefore, they are not yet capable to bridge the aforementioned scalability gap, and therefore new-generation technological platforms are highly-demanded.

We believe that our-proposed platform is one of the first attempts towards implementing such new-generation content marketing platforms, and we are foreseeing an impact of our technology on the future of digital advertisement industry once being deployed in a real-world scenario. We, therefore, see the online environment of the WSDM conference to be a perfect opportunity for such deployment which will serve as the first step towards global platform roll-out in its future development iterations. 

\vspace{-1em}
\section{Conclusion}
\vspace{-0.5em}
In this technical demonstration, we have presented the World's first personality-driven marketing content generation platform, called SoMin.ai~\cite{farseev2018somin}. The platform makes a further step towards helping digital marketers in producing quality digital marketing content at a large scale and lower cost. The experimental results demonstrate that SoMin.ai is able to achieve good personality profiling performance when being applied in the large-scale marketing content generation scenario. The generated personalized marketing content will allow marketers to boost their digital efforts by conducting efficient digital advertising campaigns easier, driven by large-scale personality data, and with no guesswork involved. Additionally, the proposed technology could be also adopted for use in various programmatic media buying platforms, such as Facebook, for achieving higher content personalization levels and better user experience.

\bibliographystyle{ACM-Reference-Format}
\balance
\vspace{-0.5em}
\bibliography{references}

\end{document}